\documentclass[aps,pra,twocolumn,amsmath,amssymb,longbibliography]{revtex4}
\usepackage{graphicx}
\usepackage{hyperref}
\usepackage{bm}
\usepackage{mathtools}
\usepackage{xcolor}
\usepackage[T1]{fontenc}
\usepackage[utf8]{inputenc}
\usepackage{changepage} 

\newcommand{\be}{\begin{equation}}\newcommand{\ee}{\end{equation}}
\newcommand{\bea}{\begin{eqnarray}}\newcommand{\eea}{\end{eqnarray}}
\newcommand{\brr}{\begin{array}}\newcommand{\err}{\end{array}}
\newcommand{\bit}{\begin{itemize}}\newcommand{\eit}{\end{itemize}}
\newcommand{\ben}{\begin{enumerate}}\newcommand{\een}{\end{enumerate}}
\newcommand{\ba}{\begin{array}}\newcommand{\ea}{\end{array}}

\definecolor{darkred}{rgb}{.8,0,0}
\definecolor{darkblue}{rgb}{0,0,0.7}

\begin{document}
\title{Variation on the theme of Jarzynski's inequality}

\author{Dani R. Castellanos}
\email{castedan@cvut.cz}
\affiliation{FNSPE,
Czech Technical University in Prague, B\v{r}ehov\'{a} 7, 115 19, Prague, Czech Republic}
\author{Petr Jizba}
\email{p.jizba@fjfi.cvut.cz}
\affiliation{FNSPE,
Czech Technical University in Prague, B\v{r}ehov\'{a} 7, 115 19, Prague, Czech Republic}
\date{\today}
%

\begin{abstract}
The Jarzynski equality, which relates equilibrium free-energy difference to an average of non-equilibrium work, plays a central role in modern non-equilibrium statistical thermodynamics. In this paper, we study a weaker consequence of this relation, known as Jarzynski's inequality, which can be formally obtained from the Jarzynski equality via Jensen's inequality. We identify and analyze several extensions of Jarzynski's inequality that go beyond its direct derivation from the Jarzynski equality. In particular, we consider chemical systems both in the linear-response regime and away from linear thermodynamics. Furthermore, by employing functional-integral techniques, we extend Jarzynski's inequality to many-body statistical systems described by quantum field theory. Salient issues, such as connections of the Jarzynski inequality with the maximum work theorem and the Landau--Lifshitz  theory of fluctuations, are also discussed.

\end{abstract}

\maketitle

\section{Introduction}

The Jarzynski equality (JE)~\cite{jarzynski1997nonequilibrium,jarzynski1997bnonequilibrium}, first introduced by Jarzynski in 1997, has become an instrumental tool in nonequilibrium statistical physics, providing a crucial link between equilibrium and nonequilibrium thermodynamics by allowing the equilibrium parameters of a system to be obtained from its nonequilibrium dynamics. JE has been verified in a wide variety of contexts, ranging from experiments on biomolecules~\cite{Liphardt,Collin}, through mesoscopic mechanical systems~\cite{Douarche,Blickle} to numerical simulations~\cite{jarzynski1997bnonequilibrium,Gupta}. There are also important generalizations of JE to the quantum regime~\cite{An,Talkner}. The statement of JE is surprisingly simple, namely
%
\begin{eqnarray}
\langle e^{- \beta W} \rangle_{\gamma} \ = \ e^{-\beta \Delta F} \;\;\; \Leftrightarrow \;\;\; \log \langle e^{-\beta W_{\rm{diss}}} \rangle_{\gamma} \ = \ 0\, .
\label{1.aa}
\end{eqnarray}
Here $\beta = T^{-1}$ is the inverse temperature (we work in natural units where $k_B \!= \! \hbar\! = \!1$), $\Delta F= F_X - F_Y$ and $W_{\rm{diss}} \!=\! W - \Delta F$ is the work dissipated during the process $X\rightsquigarrow Y$.  The brackets $\langle \cdots \rangle_{\gamma}$  denote an average over an ensemble $\gamma$ of all possible realizations of the process, which bring the system from the initial equilibrium state $X$ to a new, generally nonequilibrium state under the same external conditions as the equilibrium state $Y$. In the adiabatic limit, when the process is infinitely slow, the work $W$ done on the system in each realization is numerically equal, so the average $\langle \cdots \rangle_{\gamma}$ can be omitted and JE reduces to the thermodynamic equality $W_{\rm{diss}} = 0$.

Because ``$\log$'' in the second equality in (\ref{1.aa}) is a concave function, one can use Jensen's inequality to obtain
\begin{eqnarray}
&&-\beta \langle W_{\rm{diss}}  \rangle_{\gamma} \ \le \ \log \langle e^{-\beta W_{\rm{diss}} }\rangle_{\gamma} \ = \ 0\, ,
\end{eqnarray}
which is equivalent to 
\begin{eqnarray}
\langle W  \rangle_{\gamma} \ \ge \ \Delta F\, .
\label{2.aa}
\end{eqnarray}
We will refer to the above implication of JE as \emph{Jarzynski's inequality} (JI).
It may be noted that Jensen's inequality ensures that JI is saturated only in the case when all processes in the ensemble $\gamma$ have the same weight, i.e., when all considered process realizations are adiabatically slow, i.e., quasi-static.
Historically, the first derivation of JI --- which predated the actual formulation of JE --- was obtained for mechanical systems close to equilibrium using the fluctuation-dissipation theorem; see, for example~\cite{Fred} and a more recent account in~\cite{Hermans}.
An important feature of JI, discussed further below, is that it admits extensions beyond the regimes typically assumed in derivations of JE.
In this connection it might be stressed that the form of JI is reminiscent of the maximum work theorem (MWT)~\cite{Callen,Fermi}, which states that
$W \ge \Delta F$, where $W$ denotes the work performed by a system along an arbitrary, generally irreversible path from an initial state $X$ to a final state $Y$, provided both states are at the same temperature. However, despite this formal similarity, the mathematical foundations of MWT and JI are fundamentally different.
The MWT is a purely thermodynamic result that follows from Clausius' inequality, whereas JI is a dynamical statement derived within a microscopic Hamiltonian framework.

To test the validity of JE in the laboratory, as opposed to numerical simulations, a system under consideration must be microscopic or, at most, mesoscopic in size. Otherwise, due to large fluctuations, one would have to perform an unreasonably large number of repetitions of work measurements along state-space trajectories $\gamma$ to achieve the desired precision~\cite{jarzynski1997nonequilibrium}. As a consequence, direct experimental verification of JE in macroscopic systems remains highly challenging~\cite{Hahn,Liu,Huang}.  Since MWT is, in principle, applicable to macroscopic systems even beyond the near-equilibrium regime, it can be used as a useful benchmark for assessing the performance and scope of JI in regimes that are rarely accessible to direct tests of JE.
One of the aims of this paper is therefore to clarify the differences in applicability and scope between the MWT and JE-based JI.
A second objective of this work is to investigate several extensions of JI that are distinct from the MWT yet remain relevant for thermodynamic phenomenology. In particular, we focus on chemical systems both in the linear-response regime and away
from linear thermodynamics.
Finally, by employing functional-integral techniques, we extend JI to many-body statistical systems formulated within a quantum field–theoretical framework.




The paper is organized as follows. In the next section,  we will revisit some of the aspects of MWT that are required in the main body of the text. We also derive JE (and ensuing JI) by employing a path-integral formalism. The latter will not only provide a transparent derivation of JE, but it will also make explicit the underlying  assumptions, approximations, and domain of validity underlying of the proof, thus clarifying the scope of JE's applicability. 
In Sec.~\ref{section 3.a}, we derive JI for chemical systems operating in the linear regime by combining the entropy production inequality with Onsager's reciprocal relations and the Landau--Lifshitz fluctuation theory. 
In Sec.~\ref{chapter 4.a}, we extend JI to chemical systems that lie outside the domain of linear irreversible thermodynamics. In particular, we obtain the extension for isothermal-isobaric and isothermal-isochoric chemical processes. In Sec.~\ref{A1aa}, we employ the functional-integral approach to derive Bogoliubov--Feynman inequality in the context of statistical quantum field theory. This formulation subsequently allows us to obtain JI, hence broadening applicability of JI to quantum field-theoretical systems.
Finally, Sec.~\ref{conclusions} summarizes our results and identifies potential avenues for future research. Additional technical considerations related to Legendre transformations in irreversible processes are relegated to Appendix~\ref{A1cc}.

\section{Maximum work theorem and Jarzynski's inequality \label{Section2}}

For the sake of consistency and in order to set up our notation, we will now briefly review the key aspects of both MWT and JE that will be needed in the following sections.
In our expositions, we loosely follow Refs.~\cite{jarzynski1997nonequilibrium,jarzynski1997bnonequilibrium,Callen,Fermi,jarzynski2008nonequilibrium}.

\subsection{Maximum work theorem}

\begin{figure}[h]
\centering
\includegraphics[width=8.5cm]{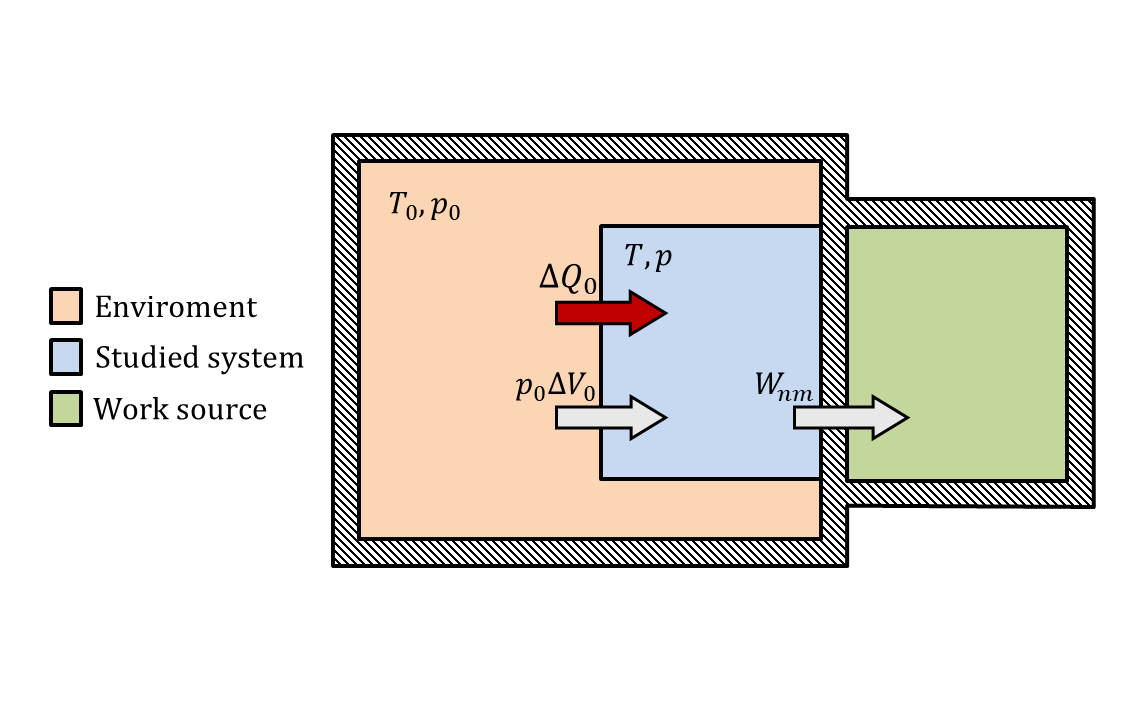}
  \caption{As an illustration of MWT, we consider an isolated system consisting of a system under study, an environment, and a work source.\label{Figure.1a}}
\end{figure}

Let us start with MWT. To do so, we consider an isolated system consisting of a system to be studied, an environment, and a work source that is adiabatically isolated from both the studied system and the environment, see Fig.~\ref{Figure.1a}. We denote the temperature and pressure of the environment as $T_0$ and $p_0$, respectively, and assume them to be constant during the process. Similarly, the temperature and pressure of the system under study are $T$ and $p$, respectively. We further assume that the studied system is not necessarily in thermal equilibrium with the environment, i.e. $T_0 \neq T$ and $p_0 \neq p$, in general.

Our goal is to find conditions under which the studied system performs maximum work on the work source. If there were no environment, the work done would directly equal the change in internal energy $\Delta U$ of the studied system. However, the existence of the environment makes the work performed on the work source ambiguous, since the studied system can exchange both work and heat with the environment.

We denote as ${{W}}_{\rm{nm}}$ the {\em non-mechanical work} that is performed by the studied system on the work source. By changing the volume $V_0$ of the environment by $\Delta V_0$, the environment will perform the work $+p_0\Delta V_0$ on the studied system. Similarly, we denote the heat received by the studied system from the environment as $-\Delta Q_0$. The negative sign reflects the fact that the heat ``delivered'' by the environment to the studied system reduces the heat of the environment by $\Delta Q_0$, hence $\Delta Q_0 < 0$.
Consequently, the change of the internal energy of the studied system is
\begin{eqnarray}
\Delta U \ = \ - \ {{W}}_{\rm{nm}} \ - \ p_0 \Delta V \ - \ \Delta Q_0\, .
\label{2.3.cc}
\end{eqnarray}
Here, we used the fact that  $V_0 + V =~$const. implies  $\Delta V_0 = -\Delta V$.

Now, we denote as $S_0$ the entropy of the environment and as $S'$ the entropy of the work source. This implies that
\begin{eqnarray}
\Delta Q_0 \ = \ T_0 \Delta S_0\, , \;\;\;\; \Delta S' \ = \ 0\, .
\end{eqnarray}
By employing Clausius' inequality
\begin{eqnarray}
\Delta\left(S \ + \ S_0 \ + \ S' \right) \ = \ \Delta S \ + \ \Delta S_0 \ + \ \Delta S' \ \ge \ 0 \, ,~~~~
\end{eqnarray}
we get
\begin{eqnarray}
- \Delta Q_0 \ \le \ T_0 \Delta S\, .
\label{2.6.gb}
\end{eqnarray}
Combining Eqs.~(\ref{2.3.cc}) and~(\ref{2.6.gb}) and the fact that $p_0$ and $T_0$ are constants gives
\begin{eqnarray}
{{W}}_{\rm{nm}} &\le& - \Delta U \ - \ p_0 \Delta V \ + \ T_0 \Delta S \nonumber \\[2mm] &=&  - \Delta\left(U \ - \ T_0 S \ + \ p_0 V  \right)\, .
\label{2.7.fg}
\end{eqnarray}
Thus, we see that the maximum work ${{W}}_{\rm{nm}}$ performed by the studied system on the work source is equal to
\begin{eqnarray}
{{W}}_{\rm{nm, \ \! max}}  \ = \ - \Delta\left(U \ - \ T_0 S \ + \ p_0 V \right)\,
.
\end{eqnarray}
This happens when Clausius' inequality is saturated, i.e. when the process is {\em reversible}.

Let us now assume that the pressure $p$ and the temperature $T$ of the studied system are the same throughout the system, but not necessarily constant over time. In other words, we assume that the process is slow enough that the studied system can internally homogenize its temperature and pressure (it is in internal thermodynamic equilibrium) at any time , but it is not necessarily in equilibrium with the environment, which is fixed at $p_0$ and $T_0$. We should emphasize that we do not assume that the system is in full thermodynamic equilibrium at any given time, in fact it might still be, e.g., in chemical non-equilibrium (at a given pressure and temperature there can still progress various chemical reactions, dissociation processes, etc.). Thus, the process is not necessarily reversible.

In particular, we note that the change in the Gibbs free energy $\Delta G$ during the process
 $X\rightsquigarrow Y$ with the realization described by the {\em state-space trajectory} $\eta$ is
 \begin{eqnarray}
 \Delta G \ &=& \  \int_{\eta} dG \ = \  \int_{\eta} dU \ - \ \int_{\eta} d(TS) \ + \ \int_{\eta} d(pV)\nonumber \\[2mm] &=& \ \Delta U \ - \ (TS)|_{Y} \ + \ \ (TS)|_X \nonumber \\[2mm]  &+& \ (pV)|_{Y} \ - \ (pV)|_X \, .
 \label{2.9cf}
 \end{eqnarray}
In the case when the studied system is in states $X$ and $Y$ in {\em thermodynamic equilibrium} with the environment (i.e. $T(t_X) = T(t_Y) = T_0$ and $p(t_X) = p(t_Y) = p_0$), we can rewrite~(\ref{2.9cf}) as
\begin{eqnarray}
\Delta G \ &=& \ \Delta U \ - \ T_0 \Delta S \ + \ p_0 \Delta V \nonumber\\[2mm]  &=& \ \Delta \left(U \ - \ T_0S \ + \ p_0V   \right)\, .
\label{2.10.cv}
\end{eqnarray}
This is true even without assuming that the studied system is in the states $X$ and $Y$ in {\em full thermodynamic equilibrium} with the environment. If we compare~(\ref{2.10.cv}) with~(\ref{2.7.fg}), we obtain
\begin{eqnarray}
{{W}}_{\rm{nm}} \ \le \ - \Delta G \, ,
\label{2.11.cf}
\end{eqnarray}
where the inequality is saturated (i.e. ${{W}}_{\rm{nm}} $ is maximal) only for reversible processes.

Inequality~(\ref{2.7.fg}) can also be recast in the form
\begin{eqnarray}
\tilde{W} \ = \ {{W}}_{\rm{nm}} \ + \ p_0 \Delta V \ \le \ - \Delta\left(U \ - \ T_0 S \right)\, .
\label{2.12.ff}
\end{eqnarray}
We might note that $-\tilde{W}  \equiv  W$ corresponds to the {\em total work} performed on the studied system.
At this stage it is convenient to  shift our focus from the description in terms of the Gibbs free energy to that of the Helmholtz free energy. In such a case, the change in the Helmholtz free energy $\Delta F$ during the process $X\rightsquigarrow Y$ realized along the generic state-space trajectory $\eta$ is
 \begin{eqnarray}
 \Delta F \ &=& \ \int_{\eta} dF \ = \   \int_{\eta} dU \ - \ \int_{\eta} d(TS) \nonumber\\[2mm] &=& \ \Delta U \ - \ (TS)|_{Y} \ + \ (TS)|_X \, .
 \label{2.13bb}
 \end{eqnarray}
In the case when the studied system is in the states $X$ and $Y$ in {\em thermal equilibrium} with the environment (i.e. when $T(t_X) = T(t_Y) = T_0$, while $p(t_X)$ and $p(t_Y)$ are arbitrary), we can rewrite~(\ref{2.13bb}) as
\begin{eqnarray}
\Delta F \ = \ \Delta U \ - \ T_0 \Delta S \ = \ \Delta \left(U \ - \ T_0S \right)\, .
\end{eqnarray}
By comparison with~(\ref{2.12.ff}) we can write
\begin{eqnarray}
\tilde{W} \ \le \ - \Delta F\, .
\label{2.15.cv}
\end{eqnarray}
The inequality is again saturated (i.e. $\tilde{W}$ is maximal) only for reversible processes.
Inequalities~(\ref{2.11.cf}) and~(\ref{2.15.cv}) constitute MWT.

Connection with JI is obtained when we realize that
the total work $W$ performed on the studied system during the process $\eta$ (having the boundary conditions $T(t_X) = T(t_Y) = T_0$) satisfies inequality
\begin{eqnarray}
W \ \ge \ \Delta F\, .
\label{2.16.cc}
\end{eqnarray}
Since $\Delta F$ is independent of the process $\eta$ (with given boundary conditions), we might average both sides of~(\ref{2.16.cc}) with respect to any distribution that is defined on the space of
all $\eta$ processes with aforementioned Dirichlet boundary conditions. This gives
%
\begin{eqnarray}
\langle W \rangle_{\eta} \ \ge \ \Delta F\, .
\label{2.17.cc}
\end{eqnarray}
Note that a choice of distribution is basically immaterial for inequality~(\ref{2.17.cc}) to hold, although for some ensembles of $\eta$ processes there may be a natural physical choice of measure (see next subsection).

Similarly, we can recast MWT~(\ref{2.11.cf})  in the form
\begin{eqnarray}
\tilde{W}_{\rm{nm}} \ \ge \ \Delta G\, ,
\label{18.cvf}
\end{eqnarray}
where $\tilde{W}_{\rm{nm}}$ denotes the non-mechanical work done by the work source on the studied system, under the condition that $T(t_X) = T(t_Y) = T_0$ and $p(t_X) = p(t_Y) = p_0$. Again, an ensemble average over all realizations of the process gives
\begin{eqnarray}
\langle \tilde{W}_{\rm{nm}}\rangle_{\eta} \ \ge \ \Delta G\, ,
\label{16.ccff}
\end{eqnarray}
which might be viewed as yet another form of JI.


\subsection{Jarzynski’s inequality}

In the previous subsection, we saw that irreversible processes are naturally characterized by inequalities, such as those derived by MWT [see Eqs.~(\ref{2.16.cc}) and~(\ref{18.cvf})]. Remarkably, when both thermal and quantum fluctuations are properly taken into account, these inequalities can, under rather general conditions, be reformulated as exact equalities. In particular, the non-equilibrium work relation~(\ref{2.16.cc}) can be recast in the form of the Jarzynski equality~(\ref{1.aa}). In the following, we establish this result using a path-integral formalism. This will not only provide a transparent derivation of the equality, but also makes explicit the assumptions, approximations, and domain of validity underlying the proof, thus clarifying the limitations of its applicability.

To this end, we consider a classical system described by generalized coordinates $q(t)$ and momenta $p(t)$, evolving under a time-dependent Hamiltonian $H(q(t), p(t), \lambda(t))$. Here $\lambda(t)$ denotes an externally controlled parameter --- such as the position of a piston or an applied electric or magnetic field --- that varies with time. The total work $W$ performed on the system during the process from $t = 0$ to $t = \tau$ is then given by
\begin{equation}
    W \ = \  \int_0^\tau dt \, \frac{\partial H(q(t), p(t), \lambda(t))}{\partial \lambda} \dot{\lambda}(t)\, .
\end{equation}
We now compute the average $\langle e^{-\beta W} \rangle_{\gamma}$ over an ensemble of all possible {\em phase-space trajectories} $z(t) = \{q(t), p(t)\}$ obtained by sampling initial conditions from a canonical ensemble (defined at a heat-bath temperature $1/\beta$) and stochastically evolving from each of these initial conditions.

We assume further that the dimension of both  $q(t)$ and $p(t)$ is $n$.
The average of the exponential  work is
\begin{eqnarray}
    \langle e^{-\beta W} \rangle_{\gamma}  &=&  \int \mathcal{D}q \mathcal{D}p \, \mathcal{P}[q, p, \lambda] \ \! e^{-\beta W[q,p,\lambda]}\nonumber
    \\[2mm]
     &=&  \int \mathcal{D}z \, \mathcal{P}[z,\lambda] \ \! e^{-\beta W[z]}\, .
     \label{2.20cc}
\end{eqnarray}
Here $\mathcal{P}[q, p, \lambda] = \mathcal{P}[z, \lambda]$ represents the probability density on the space of phase-space trajectories. An important assumption is that $\mathcal{P}[z,\lambda]$ is a (non-stationary) Markov process. Thus, the time evolution of the system along each trajectory can be considered as an adiabatic process. In the discrete time-sliced form the path integral~(\ref{2.20cc}) reads
\begin{widetext}
\begin{eqnarray}
 \langle e^{-\beta W} \rangle_{\gamma}  &=& \lim_{N \rightarrow \infty} \mathcal{N}_N \!\!\int \prod_{k=0}^N d z_k \ \! p(z_0)p_{\lambda_1}(z_1|z_0) p_{\lambda_2}(z_2|z_1)p_{\lambda_3}(z_3|z_2) \times \cdots \nonumber \\[2mm]
 &&\cdots \times p_{\lambda_{N-1}}(z_{N-1}|z_{N-2})p_{\lambda_N}(z_{N}|z_{N-1}) \ \! e^{-\beta \sum_{l=0}^N \delta W[z_l, \lambda_l]}
 \, ,
 \label{2.21.fg}
\end{eqnarray}
\end{widetext}
where $\mathcal{N}_N = 1/(2\pi \hbar)^{nN}$ is the $N$-dependent measure factor (Gibbs factor) in the phase-space path integral~\cite{kleinert}. We have also employed a short-hand notation $z_k = z(t_k)$. In addition, due to the assumed Markov property of the stochastic trajectories $z(t)$, time slicing together with
Bayes' theorem (and ensuing chain rule for conditional probability) allowed us to decompose $\mathcal{P}[z,\lambda]$ into a product of the single-time step transition probabilities and the initial-time probability.
%
%
We can now use two additional pieces of information, namely that the system is initially (i.e. at time $t_0 = 0$) in thermal equilibrium with its environment, so that the ensemble of trajectories at the initial time is distributed according to
\begin{eqnarray}
p(z_0) \ = \ \frac{e^{-\beta H(z_0)}}{Z_0}\, ,
\end{eqnarray}
and that the (mechanical) work done on the system during the process from $t_k$ to $t_{k+1}$ is given by
\begin{eqnarray}
\delta W[z_k, \lambda_k] \ = \  H(z_k,\lambda_{k+1}) \ - \  H(z_{k},\lambda_{k})\, ,
\end{eqnarray}
which is a discrete version of the continuous relation
\begin{equation}
   \delta W[q,\lambda] \ = \  \int_{t_k}^{t_{k+1}} dt \, \frac{\partial H(q(t), p(t), \lambda(t))}{\partial \lambda} \dot{\lambda}(t)\, .
\end{equation}\\
With this, we can rewrite (\ref{2.21.fg}) in the form
\begin{widetext}
\begin{eqnarray}
 \langle e^{-\beta W} \rangle_{\gamma}  = \lim_{N \rightarrow \infty} \mathcal{N}_N
 \ \!\int \prod_{l=0}^N d z_l \ \!  \frac{e^{-\beta H(z_0)}}{Z_0} \left[\prod_{k=1}^N p_{\lambda_k}(z_{k}|z_{k-1}) \ \! e^{-\beta [H(z_{k-1},\lambda_{k}) - H(z_{k-1}, \lambda_{k-1})]}  \right] .
 \label{24.cc}
\end{eqnarray}
\end{widetext}
Let us now use the fact that a marginal (or posterior) probability obtained from the transition probability $p_{\lambda_{k}}(z_k|z_{k-1})$ and a prior distribution $p_{\lambda_{k}}(z_{k-1})$ satisfy the relation
\begin{eqnarray}
p_{\lambda_k}(z_{k}) \ = \ \int \frac{dz_{k-1}}{(2\pi \hbar)^n} p_{\lambda_{k}}(z_k|z_{k-1}) \ \! p_{\lambda_{k}}(z_{k-1})\, .
\end{eqnarray}
So, by choosing as a prior distribution $p_{\lambda_{k}}(z_{k-1}) = e^{-\beta H(z_{k-1}, \lambda_{k}) }/Z_{\lambda_k}$ and by assuming that the transition probability will lead us in the time interval $\Delta t = t_k - t_{k-1}$ to the state described by the probability $p_{\lambda_{k}}(z_{k}) = e^{-\beta H(z_{k}, \lambda_{k}) }/Z_{\lambda_k}$  (which can be understood as the {\em detailed balance relation}~\cite{jarzynski1997bnonequilibrium}) one could rewrite Eq.~(\ref{24.cc}) as
\begin{widetext}
\begin{eqnarray}
\langle e^{-\beta W} \rangle_{\gamma}   &=& \frac{1}{Z_0}\lim_{N \rightarrow \infty} \mathcal{N}_2
 \ \!\int d z_N \int d z_{N-1}  \ \! e^{-\beta H(z_{N-1}, \lambda_{N-1})} \ \!  p_{\lambda_N}(z_{N}|z_{N-1})\ e^{-\beta [H(z_{N-1},\lambda_N) - H(z_{N-1}, \lambda_{N-1})]}
\nonumber \\[2mm] &=& \ \frac{1}{Z_0} \lim_{N \rightarrow \infty} \ \!\int \frac{d z_N}{(2\pi \hbar)^n}  \ \! e^{-\beta H(z_N,\lambda_N)} 
   \ = \ \frac{Z(\lambda(\tau))}{Z(\lambda(0))}
 \, ,
 \label{2.22.fg}
\end{eqnarray}
\end{widetext}
where on the first line we used the fact that $H(z_0) = H(z_0, \lambda_0)$, and on the last one line, we employed $Z_0 = Z_{\lambda_0} = Z(\lambda(0))$ and $\lim_{N \rightarrow \infty} Z_{\lambda_N} = Z(\lambda(\tau))$.

To further interpret the result~(\ref{2.22.fg}), let us recapitulate the situation.
At time $t = 0$, the system starts in thermal equilibrium with partition function
\begin{equation}
    Z(\lambda(0)) \  = \  \int \frac{dq(0) dp(0)}{(2\pi \hbar)^n} \, e^{-\beta H(q(0), p(0), \lambda(0))}\, .
\end{equation}
At time $t = \tau$, the external parameter has changed from $\lambda(0) = \lambda_0$ to $\lambda(\tau)$. If the system were allowed to equilibrate at $\lambda(\tau)$, the corresponding partition function would have the conventional Boltzmann form
\begin{equation}
    Z(\lambda(\tau)) \ = \  \int \frac{dq(\tau) dp(\tau)}{(2\pi \hbar)^n} \, e^{-\beta H(q(\tau), p(\tau), \lambda(\tau))}\, .
\end{equation}
Thus, the ratio of the partition functions is related to a free energy difference $\Delta F$ as
\begin{equation}
    e^{-\beta \Delta F} \ = \  \frac{Z(\lambda(\tau))}{Z(\lambda(0))}\, .
    \label{2.32.kl}
\end{equation}
Consequently, (\ref{2.22.fg}) together with (\ref{2.32.kl}) yields
\begin{equation}
    \langle e^{-\beta W} \rangle_{\gamma} \  =  \ e^{-\beta \Delta F}\, ,
    \label{2.32.cg}
\end{equation}
which is the sought JE. As already mentioned, JI follows from~(\ref{2.32.cg}) when we employ Jensen's inequality.

\vspace{3mm}

A few comments are now in order:

\begin{enumerate}
  \item The manner in which JE was derived shows that JE represents a {\em mathematical identity} that relates irreversible work, characterized by a time-dependent external parameter $\lambda$, to the equilibrium free energy difference~\cite{jarzynski1997nonequilibrium,jarzynski1997bnonequilibrium,jarzynski2008nonequilibrium}.
  \item Although JE is a mathematical identity, it is based on three physical assumptions; a) the Markovianity of the ensemble of trajectories described by the probability density ${\mathcal{P}[z,\lambda]}$, b) a detailed balance relation between a prior distribution $p_{\lambda_k} (z_{k-1}) = e^{-\beta H(z_{k-1},\lambda_k)}/Z_{\lambda_k}$ and a marginal distribution $p_{\lambda_k} (z_k) = e^{-\beta H(z_k,\lambda_k)}/Z_{\lambda_k}$, even if in the time interval $(t_{k-1}, t_k)$ the studied system is neither at the equilibrium temperature $1/\beta$ nor in the state described by the
  canonical distribution, and c) the studied system is allowed to equilibrate with the surrounding heat bath both at the initial time $t=0$ and at the final time $t = \tau$.
  \item If we compare JI that is obtained from MWT with that obtained from JE, we see that they generally do not carry the same information. First, the MWT-based JI is related to a broader class of studied systems, which includes systems that are not necessarily mechanical in nature. Similarly,  the work $W$ need not to be mechanical work. Second, the MWT inequality holds separately for each trajectory from the ensemble $\eta$. Recall that the $\eta$ ensemble is an ensemble of state-space trajectories that, at any given time, represent the internal thermodynamic equilibrium states of the studied system.  Consequently, the corresponding non-equilibrium state-space mean value $\langle \cdots \rangle_{\eta}$ is not easily translated to the non-equilibrium phase-space mean value $\langle \cdots \rangle_{\gamma}$. In fact, while the assumed Markovianity of the ensemble of trajectories $\gamma$ implies that the ensemble of state-space trajectories $\eta$ corresponds to systems in internal thermodynamic equilibrium, the opposite is not true, since state-space trajectories from $\eta$ might possess memory effects, e.g. via chemical non-equilibrium.
\end{enumerate}



%
%
%




\section{Chemical systems in the linear response regime \label{section 3.a}}

In this section, we show how the Jarzynski inequality follows from the entropy production inequality.
Let us first consider a network of chemical reactions as an open system, so that the evolution of the system will be out of equilibrium. We can write the infinitesimal change in the entropy of the system as
\begin{equation}
dS_s \ = \ dS_{\small{E}} \ + \ dS_i \, ,
\label{3.3.aa}
\end{equation}
where $S_s$ is the entropy of the system, $dS_{\small{E}} $ is the exchange of entropy with the environment (e.g. heat reservoir), and $dS_i$ is the production of entropy by irreversible processes inside the system (with $dS_i \geq 0$ and equality only for a reversible process). In the special case where the system exchanges heat $\delta Q$ with a reservoir at temperature $T$, we have
\begin{eqnarray}
 dS_{\small{E}} \ = \ \frac{\delta Q}{T}\, .
\end{eqnarray}
In order to extend the above thermodynamic formalism we might introduce time $t$, see e.g.~\cite{prigogine1955thermodynamics}.  In such a case, we can write that
\begin{equation}
\frac{dS_s}{dt} \ = \  \frac{dS_{\small{E}} }{dt} \ + \  \frac{dS_i}{dt}\, ,
\label{3.35.cf}
\end{equation}
where the entropy-production rate $\dot{S}_i$ is always positive or zero. i.e.
\begin{equation}
\frac{dS_i}{dt}  \ \geq \ 0\, .
\label{3.35.ff}
\end{equation}
The latter is a direct consequence of the second law of thermodynamics.

Let us now focus on the (chemical) affinity associated with chemical reactions.
The concept of chemical affinity refers to the tendency of a chemical reaction to proceed in a particular direction and it can be related to chemical potentials by the De~Donder relation~\cite{macdougall1937thermodynamic,young1938thermodynamic}
\begin{equation}
A \ = \ -\sum_{k} \gamma_k \mu_k \, , \label{eq:myequation11}
\end{equation}
where $\gamma_k$ represent the stoichiometric coefficients of the species involved and $\mu_k$ denote their respective chemical potentials.
The convention used is that the stoichiometric coefficients of the reactants are {\em negative} quantities, while those of the products are {\em positive}.
De~Donder's definition~(\ref{eq:myequation11})  basically states that the chemical affinity of a reaction is a weighted sum of the chemical potentials of the reactants and products, with the stoichiometric coefficients acting as weights.
In particular, when $A<0$ the reaction tends to move toward the products, whereas when $A>0$ the reaction tends to move toward the reactants. So, at chemical equilibrium, we have $A=0$, which means that there is no net driving force for  further reaction.
%
%
In the specific case where temperature $T$ and pressure $p$ are constants, the latter can also be understood as a consequence of the fact that the Gibbs free energy is minimal at equilibrium.
Indeed, if the only {\em non-mechanical work} is chemical work, then the change of Gibbs free energy change reads
\begin{eqnarray}
dG_{p,T} \ = \ \sum_{k} \mu_k dn_k\, .
\label{3.37.cv}
\end{eqnarray}
Here, $dn_k$ refers to the change in the number of moles which is related to the stoichiometric coefficients $\gamma_k$ via the relation $dn_k= \gamma_k d\xi$. Here $\xi$ represents the extent of reaction (also known as degree of advancement)~\cite{Callen}. This allows to recast~(\ref{3.37.cv}) into form
\begin{eqnarray}
\left(\frac{\partial G}{\partial \xi}\right)_{\!p,T} \ = \ \sum_{k} \gamma_k \mu_k \ = \ - A(\xi)\,,
\label{3.38.vv}
\end{eqnarray}
which is zero at equilibrium.

Since for constants $p$ and $T$ we have $dS_{\small{E}}  = \delta Q/T = dH/T$, where $H$ is the enthalpy of the system, we can rewrite  the entropy balance equation~(\ref{3.3.aa}) as
\begin{eqnarray}
dS_s \ &=& \ \frac{dH}{T} \ + \ dS_i\, , \nonumber \\[2mm] \Rightarrow\;\;dS_i \ &=& \ dS_s  \ - \ \frac{dH}{T} \ = \ - \ \frac{dG}{T}\, .
\end{eqnarray}
Using~(\ref{3.35.ff}),~(\ref{3.37.cv}) and~(\ref{3.38.vv}), we see that the affinity is related to the chemical reaction rate $\dot{\xi}$ via the entropy-production rate as
\begin{eqnarray}
\frac{dS_i}{dt} \ = \ \frac{A}{T}\frac{d\xi}{dt} \ \geq \ 0\, . 
\label{3.3.cfgf}
\end{eqnarray}
When the system is close to equilibrium (in linear response thermodynamics), we can assume that the chemical reactions occur near equilibrium. In this case, we may employ Onsager's reciprocal relations~\cite{onsager1931reciprocal,onsager1931reciprocal2} and rewrite equation~(\ref{3.3.cfgf}) as a product of generalized fluxes (i.e. reaction rates) and generalized driving forces (i.e. affinities), i.e.
\begin{eqnarray}
\frac{dS_i}{dt} \ = \ J X \ \geq \ 0\, , 
\label{42.cc}
\end{eqnarray}
where we have denoted $X = A/T$ and $J = d\xi/dt$.

In a case where we would be discussing a chemical network,
then $A \rightarrow  A_l$, $n_k\rightarrow n_{kl}$ and $\xi \rightarrow \xi_l$, where index  $l$ denotes the $l$-th reaction in the mechanism (e.g., the celebrated Brusselator~\cite{prigogine1978time} is a mechanism consisting of a sequence of four elementary reactions, and thus $l = 1,2,3,4$). In this way~(\ref{42.cc}) generalizes for chemical networks to
\begin{equation}
\frac{dS_i}{dt} \ = \ \sum_{n}J_n X_n \ \geq \ 0\, . \label{eq:myequation}
\end{equation}
%
For simplicity, in the following arguments we will mostly consider a global reaction mechanism, in which case we have only one $\xi$.

The entropy change due to fluctuations from equilibrium is then given by
\begin{equation}
\Delta S_i \ = \  \int_{\xi_e}^{\xi} dS_i \ = \   \int_{\xi_e}^{\xi} \frac{A}{T}d\xi\, .\label{eq:myequation}
\end{equation}
Since $A(\xi_e) = 0$, we can expand $A$ around $\xi_e$ and  retain only terms linear in the deviations $\delta\xi = \xi - \xi_e$. This gives
\begin{equation}
A \ = \ \left(\frac{\partial A}{\partial \xi}\right)_{\!e} (\xi - \xi_e)\label{eq:myequation}\, .
\end{equation}
Consequently
\begin{eqnarray}
\Delta S_i  \ &=& \  \int_{\xi_e}^{\xi} \frac{1}{T} \left(\frac{\partial A}{\partial \xi}\right)_{\!e} (\xi - \xi_e) \, d\xi \nonumber \\[2mm]
&=& \   \frac{1}{2T} \left(\frac{\partial A}{\partial\xi}\right)_{\!e} (\xi - \xi_e)^2\, .
\label{3.12cg}
\end{eqnarray}
%
%
%
This may be rewritten as
%
\begin{equation}
\Delta S_i  \ = \  - \frac{1}{2}bX^2\, ,\label{eq:myequation}
\end{equation}
where $X \equiv \delta \xi = \xi - \xi_e$ and
\begin{equation}
b \ = \ -\frac{1}{T} \left(\frac{\partial A}{\partial \xi}\right)_{\!e}\, .
\end{equation}
Note that since $S_i$ has a maximum for $\xi = \xi_e$, we have that $(\partial S_i/\partial \xi)_{e} = 0$ and $(\partial^2 S_i/\partial \xi^2)_{e} <0$, so $b >0$. This can also be seen directly from~(\ref{3.35.cf}).

At this point we can apply the Landau--Lifshitz (LL) theory of fluctuations~\cite{landau2013statistical}, in which framework, the density probability of the various values of the fluctuation is given by
\begin{equation}
p(X) \ = \ \sqrt{\frac{b}{2\pi}} e^{-\frac{1}{2}bX^2}\, .
\label{eq:myequation}
\end{equation}
The mean square fluctuation is then
\begin{equation}
\langle X^2 \rangle \ = \ \int_{-\infty}^{\infty} p(X)X^2 \, dX \ = \  -{T}/{\left(\frac{\partial A}{\partial \xi}\right)_{\!e}}\, .
\end{equation}
%
%
%
The scope of the LL theory of fluctuations is only for such times $\tau$, which characterize the rate of change of $\xi$, for which the condition $T \gg \hbar/\tau$ holds~\cite{landau2013statistical}. In particular, if changes of $\xi$ are too fast (i.e. $\tau$ is too small), such fluctuations cannot be treated thermodynamically.

In the case when the number of reactions in the mechanism would be greater than one, the corresponding $\Delta S_i$ would be given by
\begin{equation}
\Delta S_i \ = \  \int_{\mathcal{C}} \sum_{k} \frac{A_k}{T} d\xi_k\, ,
\label{51.bh}
\end{equation}
where $A_i(\boldsymbol{\xi}) $ is the affinity of the $i$-th reaction, $\xi_i$ are the degrees of advancement (forming a $\boldsymbol{\xi} $-vector), and
$\mathcal{C}$ is a curve connecting equilibrium $\boldsymbol{\xi}_e$ to the final state $\boldsymbol{\xi}$.


By expanding $A_i(\boldsymbol{\xi})$ around $\boldsymbol{\xi}_e$ we get
\begin{eqnarray}
A_i(\boldsymbol{\xi}) \ &=& \  A_i(\boldsymbol{\xi}_e) \ + \  \sum_j \ \!\left( \frac{\partial A_i}{\partial \xi_j} \right)_{\!\!e} (\xi_j - \xi_{j,e})
\nonumber \\[2mm] &&+ \
\mathcal{O}((\delta \boldsymbol{\xi})^2)\, .
\end{eqnarray}
Since equilibrium is defined by $A_i(\boldsymbol{\xi}_e) = 0$, the first term vanishes and up to a linear order in $\delta \boldsymbol{\xi}$, we can write~(\ref{51.bh}) as
\begin{equation}
\Delta S_i \ = \  \frac{1}{T} \int_{\mathcal{C}}  \sum_{k,j} \ \!\left( \frac{\partial A_k}{\partial \xi_j} \right)_{\!\!e} \ \! (\xi_j - \xi_{j,e})
  d\xi_k\, .
  \label{53.klm}
\end{equation}\\
We now invoke the fact that in the linear response regime we can use a linear parameterization for the path $\mathcal{C}$ in the entropy integral~(\ref{53.klm}). This will simplify the following calculations while still capturing the essential physics of small deviations from equilibrium. To do this, we assume a smooth parameterized path from $\boldsymbol{\xi}_e$ to $\boldsymbol{\xi}$, given by
\begin{equation}
\xi_k(\lambda) \ = \  \xi_{k,e} \ + \  \lambda \delta \xi_k \quad \;\text{with}\;\;\;\; \lambda \in [0,1]\, .
\end{equation}
So that $d\xi_k = \delta \xi_k d\lambda$, and~(\ref{53.klm}) can be rewritten as
\begin{widetext}
\begin{eqnarray}
\Delta S_i &=&  \frac{1}{T}\int_0^1   \sum_{k,j} \left( \frac{\partial A_k}{\partial \xi_j} \right)_{\!\!e} [\xi_j(\lambda) - \xi_{j,e}] \ \!
\delta \xi_k \ \! d\lambda \ = \ \frac{1}{T} \sum_{i,j} \left( \frac{\partial A_k}{\partial \xi_j} \right)_{\!e} \delta \xi_k \delta \xi_j \int_0^1 \lambda d\lambda\ = \  \frac{1}{2T} \sum_{k,j} \left( \frac{\partial A_k}{\partial \xi_j} \right)_{\!\!e} \delta \xi_k \delta \xi_j \, .~~~
\end{eqnarray}
\end{widetext}
Note that in the linear response regime the matrix
\begin{equation}
\left( \frac{\partial A_k}{\partial \xi_j} \right)_{\!\!e} \, ,
\end{equation}
must be symmetric in $k$ and $j$ indices. This can be seen, for example, by noting that in order for $\Delta S_i$ to be $\mathcal{C}$ independent, then the one-form
\begin{eqnarray}
\omega_1(\boldsymbol{\xi}) \ = \ \sum_k \left[\sum_j \left( \frac{\partial A_k}{\partial \xi_j} \right)_{\!\!e}
 \delta \xi_j\right] d\xi_k\, ,
\end{eqnarray}
must be {\em exact}. The integrability condition guarantees that this is the case, provided
\begin{equation}
\left( \frac{\partial A_k}{\partial \xi_j} \right)_{\!\!e}  \ = \ \left( \frac{\partial A_j}{\partial \xi_k} \right)_{\!\!e} \, .
\end{equation}
In addition, since $S_i$ has a maximum for $\boldsymbol{\xi} = \boldsymbol{\xi}_e$, we have that the corresponding Hessian of $S_i$, i.e.
\begin{equation}
\frac{1}{T}\left( \frac{\partial A_k}{\partial \xi_j} \right)_{\!\!e} \, ,
\end{equation}
must be {\em negative definite} matrix. Consequently, the ensuing generalization of the probability density function~(\ref{eq:myequation}) would read
\begin{eqnarray}
&&\mbox{\hspace{-5mm}}p(X_1, \ldots, X_l) \nonumber \\[2mm] &&= \  \sqrt{\frac{\det(b_{ik})}{(2\pi)^l}}\exp\left(-\frac{1}{2}\sum_{i,k}^l b_{ik} \ \!  X_iX_k \right)\, ,
\end{eqnarray}
where $X_i \equiv \delta\xi_i$ and
\begin{eqnarray}
b_{ik} \ = \ -\frac{1}{T} \left(\frac{\partial A_i}{\partial \xi_k}\right)_{\!e} \ =\ \frac{1}{T} \left(\frac{\partial^2 G}{\partial \xi_i \partial \xi_k} \right)_{p,T} \, ,
\end{eqnarray}
with the covariance function
\begin{eqnarray}
\langle X_i X_k \rangle \ = \ (b^{-1})_{ik} \, .
\end{eqnarray}
Let us now assume that $\mathcal{C}$ is a curve connecting the initial equilibrium state $\boldsymbol{\xi}_e^{(1)}$ with some final equilibrium state $\boldsymbol{\xi}_e^{(2)}$. The states on the curve are not necessarily the equilibrium states, but they deviate from the equilibrium curve (with the same initial and final states) only in such a way that they do not go beyond the validity of the LL theory of fluctuations.  By using the fact that [cf. Eq.~(\ref{A.17.hj})]
\begin{equation}
dG_{p,T}  \ \leq \ \delta W_{\rm{nm}} \, ,
\end{equation}
alongside with the linear thermodynamic framework, we may write
\begin{eqnarray}
\langle\int_{\mathcal{C}}  dG_{p,T} \rangle \ \leq \ \langle \int_{\mathcal{C}} \delta W_{\rm{nm}} \rangle\, .
\label{64.nm}
\end{eqnarray}
The LHS can further be rewritten as
\begin{widetext}
\begin{eqnarray}
\langle\int_{\mathcal{C}}  dG_{p,T}) \rangle  &=&  \int_{\mathcal{C}}  d\langle G_{p,T} \rangle
\ = \ \int_{\mathcal{C}}  d  \left[G_{p,T}(\boldsymbol{\xi}_e) \ + \  \frac{1}{2} \left(\frac{\partial^2 G_{p,T}}{\partial \xi_i \xi_j}\right)_{\!\!e} \langle X_i X_j \rangle \right]\nonumber \\[2mm]
&=& \int_{\mathcal{C}}  d  \left[G_{p,T}(\boldsymbol{\xi}_e) \ + \ \frac{Tl}{2}\right] \ = \ G_{p,T}(\boldsymbol{\xi}_e^{(2)}) \ - \ G_{p,T}(\boldsymbol{\xi}_e^{(1)})  \ = \ \Delta G_{p,T}\, ,~~~~~~
\end{eqnarray}
\end{widetext}
where $l$ is a number of reaction mechanisms.  Note that we had to expand $G_{p,T}$ to the second order, as this is equivalent to the first order expansion of the affinity $A$, we used above.
On the other hand, the RHS of (\ref{64.nm}) is the averaged non-mechanical work done on the studied system along the trajectory $\mathcal{C}$. Thus we might finally write
\begin{eqnarray}
\Delta G_{p,T} \ \leq \ \langle W_{\rm{nm}}\rangle_{_{\mathcal{C}}}\, .
\label{66.lk}
\end{eqnarray}


A similar argument can be now applied to the Helmholtz free energy. In fact, from Eq.~(\ref{A.16.bv}), we have that
\begin{equation}
dF_T \  \leq \  \delta W \ = \ -pdV \ + \ \delta W_{\rm{nm}}\,
\end{equation}
By employing again assumptions of the linear thermodynamics, we  arrive at the inequality
\begin{eqnarray}
\Delta F_T  \ \leq \ \langle {W} \rangle_{\mathcal{C}}\, ,
\label{68.fg}
\end{eqnarray}
where $\langle {W} \rangle_{\mathcal{C}}$ denotes the averaged total work performed on the studied system along the trajectory $\mathcal{C}$. This is again formally identical with JI.

\vspace{3mm}

Results~(\ref{66.lk}) and~(\ref{68.fg}) deserve a few comments:

\begin{enumerate}
\item
In this section we used the entropy production inequality rather than MWT. The MWT-based JI is
related to a broader class of studied systems, which includes systems that are not necessarily chemical in nature but might have other forms of non-mechanical work. While both MWT and the entropy production inequality are tools from non-equilibrium thermodynamics, they have slightly different scopes. Particularly, in our context, we have assumed evolution with small fluctuations around an equilibrium configuration. This is pretty close, but not exactly the same as a slow evolution that was assumed in the derivation of MWT. In particular, the characteristic time of fluctuations is $\tau \gg \hbar/T$, while in MWT a typical time change is associated with {\em relaxation time}. These two time scales may not be the same in general~\cite{landau2013statistical}. In addition, while in MWT we did not have any specific distribution on the space of state-space trajectories $\eta$ (condition  given boundary conditions), in the present case the LL theory of fluctuations naturally provides a distribution.

\item In contrast to JI derived from JE~(\ref{2.32.cg}), the above results relate to a specific class of systems, namely systems close to equilibrium, where both linear thermodynamics and the LL theory of fluctuations are valid approximations.
At the same time, results~(\ref{66.lk}) and~(\ref{68.fg}) mainly address systems that are not mechanical in nature. Similarly, the work $W$ includes also a chemical work. In this respect the results go beyond JI derived from JE~(\ref{2.32.cg}).
\end{enumerate}

\section{Chemical systems beyond linear regime \label{chapter 4.a}}
\subsection{General ideas}

Let us now discuss a possible generalization of JI to systems that are far from equilibrium. Our particular attention will again be given to chemical systems.  Consider, for instance, the specific dimerization of $NO_2$, such as
\begin{eqnarray}
2NO_{2(g)} \ \rightleftharpoons  \   N_{2}O_{4(g)}\, ,
\label{IV.70.cc}
\end{eqnarray}
where $(g)$ refers to gas states. This equation states that we will have certain amount of the reactant and we will obtain certain amount of the products that is reflected in the yield of the reaction. It establishes a directional character of the process, in which we know {\em a priori} that the reaction will occur, but we don't know what yield will be obtained under the given conditions of temperature, pressure, pH, etc..

Let us now concentrate on the Gibbs free energy of a generic non-equilibrium chemical system.  Obviously, $G$ is typically defined for systems that are at or near equilibrium (linear thermodynamics). For systems that are far from equilibrium, the situation becomes more complex and the standard definition of $G$ often does not apply. In particular, far from equilibrium,  $G$ no longer provides an universal criterion for the direction of the chemical processes; instead, the dynamics is governed by entropy production and chemical affinities. Yet, the Legendre-transformed potentials, such as $G$, are not only confined to equilibrium or to the linear response domain. Rather, they continue to be valid beyond linear thermodynamics provided that local equilibrium holds, relevant time scales remain well separated, and the fundamental thermodynamic potentials retain their convexity. Under these assumptions, nonlinear effects modify only the constitutive thermodynamic relations but leave the underlying Legendre structure intact \cite{prigogine1955thermodynamics,GM}. 
%
%
In this regime, the Gibbs free energy continues to quantify the system's ability to perform work or deliver useful energy at a fixed temperature and pressure.   In the context of a chemical reaction, a negative change in Gibbs free energy, $\Delta G < 0$, indicates that the reaction can proceed spontaneously in the direction that increases the yield of the products. When $\Delta G = 0$, the reaction has reached equilibrium, and no further net change in the concentrations of reactants and products occurs; the yield is considered maximized at equilibrium. If $\Delta G > 0$, it means that the yield under given conditions can not be realized.

For fixed temperature and pressure, the total differential of $G$ reads
\begin{equation}
dG_{T,p}\ = \  \left(\frac{\partial G}{\partial \xi}\right)_{T,p} d\xi\, .
\label{4.22}
\end{equation}
From~ this, we can infer that
\begin{equation}
\frac{dG_{T,p}}{dt} \ = \  \left(\frac{\partial G}{\partial \xi}\right)_{T,p} \frac{d\xi}{dt}\, .
\label{4.23}
\end{equation}
Equation~(\ref{4.23}) allows us to evaluate $\Delta G$ through the rate of the reaction and affinity.  In the afforestated regime, we can assume the validity of the Legendre transformation~(\ref{A.6.cc}) and write
\begin{equation}
G \ = \  U \ + \  pV \ - \ TS_s\, ,
\end{equation}
and similarly, from~(\ref{A.5.hj}), we have  
\begin{equation}
G \ = \ F \ + \ pV\, .
\label{IV.74.kl}
\end{equation}
%

\subsection{Jarzynski's inequality for isothermal-isobaric chemical processes}
%
As mentioned, for chemical reactions the Gibbs free energy characterizes  the extent of the reaction. From~(\ref{IV.74.kl}), we can rewrite~(\ref{4.22}) as
\begin{equation}
dG_{T,p}(\xi) \ = \ \left(\frac{\partial F}{\partial \xi}\right)_{T,p} d\xi \ + \  p\left(\frac{\partial V}{\partial \xi}\right)_{T,p} d\xi\, .
\end{equation}
By integrating from some $\bar{\xi}$ to the equilibrium state $\xi_e$, we may write
\begin{eqnarray}
&&\int^{\xi_e}_{\bar{\xi}} \left(\frac{\partial G}{\partial \xi}\right)_{T,p} d\xi \nonumber \\[2mm] &&~~~= \  \int^{\xi_e}_{\bar{\xi}} \left(\frac{\partial F}{\partial \xi}\right)_{T,p} d\xi \ + \ p\int^{\xi_e}_{\bar{\xi}} \left(\frac{\partial V}{\partial \xi}\right)_{T,p} d\xi\, .~~~~~~
\label{IV.76.fm}
\end{eqnarray}
At this stage, we employ the inequality~(\ref{A.9.kl}) and rewrite~(\ref{IV.76.fm}) in the form
\begin{eqnarray}
&&\int^{\xi_e}_{\bar{\xi}} \left(\frac{\partial F}{\partial \xi}\right)_{T,p} d\xi\nonumber \\[2mm] &&~~~ \ \leq \ - p\int^{\xi_e}_{\bar{\xi}} \left(\frac{\partial V}{\partial \xi}\right)_{T,p} d\xi \ - \ \int^{\xi_e}_{\bar{\xi}} A(\xi) \ \! d\xi\, .~~~~~
\label{IV.77.kl}
\end{eqnarray}
For $A>0$, i.e., for reactions that proceed in the direction that increases the yield, the former inequality can be converted into a simpler form
\begin{eqnarray}
\int^{\xi_e}_{\bar{\xi}} \left(\frac{\partial F}{\partial \xi}\right)_{T,p} d\xi \ \leq \ - p\int^{\xi_e}_{\bar{\xi}} \left(\frac{\partial V}{\partial \xi}\right)_{T,p} d\xi \, .
\label{IV.78.rt}
\end{eqnarray}
We could alternatively arrive at~(\ref{IV.78.rt}) by starting from~(\ref{IV.76.fm}) and employing the inequality $\Delta G \leq 0$.

By choosing $\bar{\xi}$ to correspond to another equilibrium state --- for example, for the dimerization process~(\ref{IV.70.cc}), we assume that the system starts from an initial equilibrium state consisting solely of the reactant $2\,\mathrm{NO}_{2(g)}$ and evolves to a final equilibrium state characterized by a mixture of $\mathrm{NO}_{2(g)}$ and $\mathrm{N}_2\mathrm{O}_{4(g)}$ ---
the corresponding integrals taken along the reaction coordinate lead to the inequality
\begin{eqnarray}
\Delta F_{T,p} \ \leq \ W_{T,p}\, ,
\end{eqnarray}
where $W_{T,p}$ is the mechanical work performed on the system while going from the state $\{T,p, \xi_e^{(1)}\}$ to $\{T,p, \xi_e^{(2)}\}$.

Let us now consider directly inequality~(\ref{IV.77.kl}). In this case   
the term $-A(\xi) \ \! d\xi$ (or more generally $- \sum_r {A}_r(\boldsymbol{\xi}) \, d\xi_r$) represents the chemical work performed  on the system during an infinitesimal advancement of the reactions. In particular
\begin{eqnarray}
\delta W_{\text{chem}} \ =\  -A(\xi) \ \! d\xi\,  ,
\label{IV.80.sd}
\end{eqnarray}
or
\begin{eqnarray}
\delta W_{\text{chem}} \ =\  - \sum_r {A}_r(\boldsymbol{\xi}) \, d\xi_r\,  .
\label{IV.81.sd}
\end{eqnarray}
It should be stressed that relations~(\ref{IV.80.sd})-(\ref{IV.81.sd}) remain valid out of equilibrium (even far-from-equilibrium) as a kinematic definitions of chemical work, provided chemical potentials and reaction extents are well defined~\cite{prigogine1955thermodynamics,GM}. 

At this stage, we may define chemical work along a single reaction trajectory as
\begin{eqnarray}
W_{\text{chem}, T,p}[\xi] \ = \ - \int_0^\tau A(\xi(t)) \ \! \frac{d\xi(t)}{dt} dt\, ,
\end{eqnarray}
or
\begin{eqnarray}
W_{\text{chem},T,p}[\boldsymbol{\xi}] \ = \ - \int_0^\tau \sum_r {A}_r(\boldsymbol{\xi}(t)) \, \frac{d\xi_r(t)}{dt} dt\, .
\end{eqnarray}
With this, inequality~(\ref{IV.77.kl})
acquires the form
\begin{eqnarray}
\Delta F_{T,p} \ \leq \ W_{T,p}[\xi]\, ,
\label{IV.84.hj}
\end{eqnarray}
Here $W_{T,p}[\xi]$ denotes the total work (mechanical $+$ chemical) performed on the system  along the reaction trajectory $\xi$ connecting two equilibrium states at fixed temperature and pressure.

If ${\mathcal{P}}[\xi]$ is  a probability density on the space of reaction trajectories for isothermal-isobaric
chemical processes --- say $\bar{\gamma}$, the average total work can be formally defined through a functional integral as
\begin{eqnarray}
\langle W\ \!\rangle_{\bar{\gamma}} \ = \ \int {\mathcal{D}} \xi  \  {\mathcal{P}}[\xi] \  W_{T,p}[\xi]\, .
\label{IV.85.fg}
\end{eqnarray}
With this, inequality~(\ref{IV.84.hj}) can be rewritten in a form that is independent of the particular choice of reaction path, namely
\begin{eqnarray}
\Delta F_{T,p} \ \leq \ \langle W\ \!\rangle_{\bar{\gamma}}\, ,
\label{IV.80.ck}
\end{eqnarray}
where $\bar{\gamma}$ denotes the ensemble of reaction trajectories corresponding to isothermal-isobaric chemical processes that connect two fixed equilibrium states at times $t=0$ and $t=\tau$.

\subsection{Jarzynski's inequality for isothermal-isochoric chemical processes}

Let us now consider a chemical system evolving at constant temperature and volume. In this case, we have  [see Eq.~(\ref{A.16.bv})] that
\begin{eqnarray}
dF_{T,V} \ \leq \ \delta W_{\rm{nm}} \ = \ \delta W_{\text{chem}, T,V}\, .
\label{IV.87.kl}
\end{eqnarray}
The corresponding chemical work, $\delta W_{\mathrm{chem},T,V}$, can be expressed in terms of the chemical affinities associated with isothermal–isochoric processes through relations analogous to~(\ref{IV.80.sd}) and~(\ref{IV.81.sd}). By integrating~(\ref{IV.87.kl}) from $\bar{\xi}$ to $\xi_e$, we get
\begin{equation}
\int^{\xi_e}_{\bar{\xi}}\left(\frac{\partial F}{\partial \xi}\right)_{T,V} d\xi \ \leq \  -\int^{\xi_e}_{\bar{\xi}} A(\xi) \  \!d\xi \, .
\label{IV.88.io}
\end{equation}
The chemical work along a single reaction trajectory can again be defined as
\begin{eqnarray}
W_{\text{chem}, T,V}[\xi] \ = \ - \int_0^\tau A(\xi(t)) \ \! \frac{d\xi(t)}{dt} dt\, ,
\label{IV.89.ty}
\end{eqnarray}
or
\begin{eqnarray}
W_{\text{chem},T,V}[\boldsymbol{\xi}] \ = \ - \int_0^\tau \sum_r {A}_r(\boldsymbol{\xi}(t)) \, \frac{d\xi_r(t)}{dt} dt\, ,
\end{eqnarray}
where, in contrast to~(\ref{IV.80.sd})-(\ref{IV.81.sd}), $A(\xi)$ (or ${A}_r(\boldsymbol{\xi})$) is now chemical affinity  for isothermal–isochoric chemical processes. Eqs.~(\ref{IV.88.io}) and~(\ref{IV.89.ty}) again imply a single-reaction trajectory inequality
\begin{equation}
\Delta F_{T,V} \ \leq \ W_{\text{chem},T,V}[{\xi}]\, ,
\end{equation}
By following the same steps as in the preceding subsection, the inequality~(\ref{IV.84.hj}) can be reformulated in terms of the average chemical work as
\begin{equation}
\Delta F_{T,V} \ \leq \ \langle W_{\text{chem}} \rangle_{\bar{\gamma}}\, .
\label{IV.86.kl}
\end{equation}
Here $\langle \cdots \rangle_{\bar{\gamma}}$ denotes the average over ensemble of reaction pathways for  isothermal-isochoric chemical processes that connect two fixed equilibrium states at times $t=0$ and $t=\tau$.

\vspace{3mm}

Results~(\ref{IV.80.ck}) and~(\ref{IV.86.kl}) deserve a few comments:

\begin{enumerate}
\item The approach presented here applies to non-equilibrium regimes that extend beyond simple linear thermodynamics. This differs from MWT, which cannot be applied directly to such processes because it relies on reversibility, whereas the processes considered here typically involve dissipation and entropy production. Similarly as in MWT, inequalities could be formulated separately for individual trajectories within the ensemble $\bar{\gamma}$. However, ensembles $\eta$ and $\bar{\gamma}$ are structurally very different.

\item In contrast to JI derived from JE, see Eq.~(\ref{2.32.cg}), we explicitly consider non-mechanical systems, namely chemical systems and ensuing chemical works. An  important distinction is that the probability density ${\mathcal{P}}[\xi]$  is not necessarily related to (non-stationary) Markov process.  In fact, since reaction trajectories from $\bar{\gamma}$ might possess weak memory effects via chemical non-equilibrium (diffusion limitations, slow environmental relaxation, etc.). In this respect the path integral measure employed in~(\ref{IV.85.fg}) is just a formal measure on the space trajectories from $\bar{\gamma}$, as the path-integral measure is guaranteed only for Markov processes by the Daniell--Kolmogorov extension theorem~\cite{Burrill} and by the Chapman--Kolmogorov equation~\cite{JK}.  
Other path measures, such as the finite-time sliced measure~\cite{kleinert}, would be more appropriate. 
It should also be noted that the structure of the paths in the ensemble $ \bar{\gamma}$ is different than in $\gamma$ ensemble (reaction trajectories versus phase-space trajectories).

\item In the case of chemical systems [such as the above dimerization reaction~(\ref{IV.70.cc})]  the externally controlled parameter $\lambda(t)$ typically corresponds to some time dependent steering mechanism (light/photon control,  mechanical steering or catalytic modulation).

\end{enumerate}

\section{Bogoliubov--Feynman inequality and JI\label{A1aa}}

We finally derive JI from the Bogoliubov--Feynman (BF) inequality. This will enable us to extend the domain of applicability of JI to statistical quantum field theoretical (QFT) systems. To this end, we first establish the BF inequality using a functional-integral approach, which yields a particularly direct and efficient derivation. This method is well suited to the QFT setting and is substantially simpler than conventional operator-based treatments (see, for example, Ref.~\cite{Feynman}).

%
%
%

We start by considering a QFT system evolving in the time interval $[0,T]$ inside the heat bath of temperature $1/\beta$. We only assume that the system is in thermodynamic equilibrium at the beginning and at the end of its time evolution, but is not necessarily in equilibrium with the environment in between.  Let the corresponding initial and final time partition functions be $Z_0$ and $Z$ respectively.
The two Helmholtz free energies are connected to the Euclidean functional integral by the relation~\cite{kleinert,Calzeta}:
\begin{widetext}
\begin{eqnarray}
&&e^{-\beta F_0} \ = \  Z_0 \ = \ \!\!\underset{\beta-{\rm
periodic}}{\int}\!\!\!\!\!\!{\mathcal{D}}\phi {\mathcal{D}}\pi \ \!
\exp\left(\int_{0}^{\beta}
d \tau [i \dot{\phi}\pi - H_0(\phi,\pi)]  \right)\, ,
\label{app.1.1aa}
\\[2mm]
&&e^{-\beta F} \ = \ Z \ = \  \!\! \underset{\beta-{\rm
periodic}}{\int}\!\!\!\!\!\!{\mathcal{D}}\phi {\mathcal{D}}\pi \ \!
\exp\left(\int_{0}^{\beta} d \tau [i \dot{\phi}\pi - H(\phi,\pi)]
\right)\, .\label{app.1.1a}
\end{eqnarray}
\end{widetext}
Here the $\beta$-periodicity condition, the so-called Kubo--Martin--Schwinger (KMS) condition, is applied only to the field $\phi$ (not to its conjugate $\pi$), as
\begin{eqnarray}\label{statistical-physics-boundary-conditions}
    &&\phi (\boldsymbol{r}, \beta) \ = \ \pm \phi (\boldsymbol{r}, 0)\, ,
\end{eqnarray}
where minus sign refers to fermi fields. In both~(\ref{app.1.1aa}) and~(\ref{app.1.1a}) we have suppressed spatial integration, which is not important for our subsequent reasoning. In the following, such a suppression will also be employed.  In the case of a grand canonical ensemble, the Hamiltonian can also contain parts with chemical potentials (see comment at the end of this section).

In our argument we use QFT (to show applicability of the Bogoliubov inequality also for large quantum systems), although the passage to a few-particle picture can be obtained simply by replacing $\phi$ and $\pi$ with the corresponding position and momentum vectors and the KMS condition with the path-periodicity condition~\cite{kleinert}.

Now the expectation value of
\begin{eqnarray}
\exp\left(-\int_0^\beta d\tau \ \! (H - H_0) \right)\, ,
\end{eqnarray}
with respect to the {\em initial-time} equilibrium distribution is
\begin{widetext}
\begin{eqnarray}
\mbox{\hspace{-5mm}}\left\langle \exp\left(-\int_0^\beta d\tau \ \! (H - H_0) \right) \right\rangle_0
&=& e^{\beta F_0}
\! \!\!\!\!\!\!\underset{\beta-{\rm
periodic}}{\int}\!\!\!\!\!\!{\mathcal{D}}\phi {\mathcal{D}}\pi \
\! \exp\left(\int_{0}^{\beta} \! d \tau [i \dot{\phi}\pi -
H(\phi,\pi)] \right) \
= \ e^{\beta F_0}\ \! e^{- \beta F}.
\end{eqnarray}
\end{widetext}
%
%
In the derivation we have assumed that the functional measures ${\mathcal{D}}\phi {\mathcal{D}}\pi $ are the same for both the $H_0$ and the $H$ systems. In general, this is not the case if, for example, the two dynamics have different degrees of freedom or if they have different gauge symmetries. In such cases, the inequality cannot be proved with the present method. In fact, it can be argued that in such cases the inequality does not hold in general.

In the next step, we use Jensen--Peierls  inequality~\cite{kleinert}
%
\begin{eqnarray}
\langle e^{-x} \rangle \ \geq \ e^{- \langle x \rangle}\, ,\label{app.1.3a}
\end{eqnarray}
(valid for any kind of mean value of a random variable $x$) which allows to write
\begin{widetext}
\begin{eqnarray}
e^{\beta F_0}\ \! e^{- \beta F} \ = \ \left\langle \exp\left(-\int_0^\beta d\tau \ \! (H - H_0) \right) \right\rangle_0  \ \geq \ \exp\left(-\int_0^\beta d\tau \ \! \langle H - H_0\rangle_0 \right)\, .
\end{eqnarray}
\end{widetext}
Taking logarithm on both sides, and using the fact that $\log(\ldots)$ is a monotonic
function of its argument, we finally obtain
\begin{eqnarray}
F \ \leq \ F_0  \ + \  \frac{1}{\beta} \int_0^\beta d\tau \ \! \langle H(\tau) - H_0(\tau) \rangle_0 \, .
\label{A8.cv}
\end{eqnarray}
To better understand the structure of the second term on the RHS, we rewrite~(\ref{app.1.1aa}) and~(\ref{app.1.1a}) equivalently in operatorial language as
\begin{eqnarray}
Z_0 &=& \mbox{Tr} \left( e^{-\beta \hat{H}_0}\right),\label{A9.cc} \\[2mm]
Z&=& \mbox{Tr} \left( e^{-\beta \hat{H}}\right) \ = \ \mbox{Tr} \left( e^{-\beta \hat{H}_0} e^{\beta \hat{H}_0} e^{-\beta \hat{H}}\right) \nonumber \\[2mm] &=&  \mbox{Tr} \left( e^{-\beta \hat{H}_0}  \hat{\Lambda}(-i\beta) \right)\nonumber \\[2mm]
&=& Z_0 \left
\langle {\hat{\Lambda}}(-i\beta)  \right\rangle_0\, .
\label{A.10.cf}
\end{eqnarray}
It can be easily checked that $ \hat{\Lambda}(-i\beta)$ satisfies the differential equation
\begin{eqnarray}
\frac{d  \hat{\Lambda}(-i\beta)}{d \beta} \ = \ -\hat{H}_I(\beta)  \hat{\Lambda}(-i\beta)\, ,
\label{A10.cc}
\end{eqnarray}
with $\hat{\Lambda}(0) = 1$. Here
\begin{eqnarray}
\hat{H}_I(\beta) \ &=& \ e^{\beta\hat{H}_0} \hat{H} \ \!e^{-\beta \hat{H}_0} \ - \ \hat{H}_0 \nonumber \\[2mm] &=& \ \hat{\bar{H}} \ - \ \hat{H}_0 \, .
\end{eqnarray}
Eq.~(\ref{A10.cc}) might be solved in terms of the Dyson series  as
\begin{eqnarray}
\hat{\Lambda}(-i\beta) \ = \ T\left[\exp\left(- \int_{0}^{\beta} d \tau \ \!   \hat{H}_I(\tau) \right)  \right]\, .
\end{eqnarray}\\
Here $T[\cdots]$ denotes a \emph{temperature-ordering symbol}, which orders operators so that those with larger values of the inverse temperature parameter~$\beta$ are positioned further to the left.
The operator~$\hat{\bar{H}}$ plays a role analogous to that of the Hamiltonian~$\hat{H}$ in the Dirac (interaction) picture of conventional quantum mechanics.
Note that the operator $\hat{\bar{H}}$ (and hence also $\hat{H}_I$) evolves in the $\beta$ variable according to Hamiltonian $\hat{H}_0$. Indeed
\begin{eqnarray}
\frac{d \hat{\bar{H}} }{d\beta} \ = \ \left[\hat{\bar{H}},\hat{H}_0\right]\, ,
\end{eqnarray}
which represents conventional quantum mechanical evolution in the imaginary time $t = -i \beta$. In particular
\begin{eqnarray}
\hat{\bar{H}}(\tau_1 + \tau_2) \ &=& \   e^{-\tau_1 \hat{H}_0}  \hat{\bar{H}}(\tau_2)  e^{\tau_1 \hat{H}_0} \nonumber \\[2mm] &=& \ e^{-\tau_2 \hat{H}_0}  \hat{\bar{H}}(\tau_1)  e^{\tau_2 \hat{H}_0}\, .
\label{A.15.cf}
\end{eqnarray}
In the following, we denote all operators that ``evolve'' with respect to $\hat{H}_0$ according to the rule~(\ref{A.15.cf}) with a hat-bar symbol over the observable, e.g. $\hat{\bar{X}}$ (except $\hat{H}_0$ itself, which is time independent).

We now employ the following identity (see, e.g., Ref.~\cite{Calzeta})
\begin{widetext}
\begin{eqnarray}
&&\mbox{\hspace{-15mm}}\underset{\beta-{\rm
periodic}}{\int}\!\!\!\!\!\!{\mathcal{D}}\phi {\mathcal{D}}\pi \
\! \exp\left(\int_{0}^{\beta} \! d \tau [i \dot{\phi}\pi -
H_0(\phi,\pi) - J H_{I}(\phi,\pi)] \right)  \ = \
 \mbox{Tr}\left\{ e^{-\beta \hat{H}_0} \ \!T\left[ \exp\left(- \int_{0}^{\beta} d \tau \ \!   J(\tau)\hat{H}_I(\tau) \right)  \right]\right\}\, .
 \label{A.16.kl}
\end{eqnarray}
\end{widetext}
Here $J(\tau)$ is a source function (not an operator). Note that for $J(\tau) = 1$ we have the original relation for the partition function $Z$. By performing a functional derivation $-\delta/ \delta J(\tau)$ on both sides of the equation~(\ref{A.16.kl}) and setting $J=0$ at the end, we get
%
%
\begin{widetext}
\begin{eqnarray}
&&\mbox{\hspace{-15mm}}\langle H(\tau) - H_0(\tau)  \rangle_0
 \ = \    \!\!\!\!\!\!\!\!\!
 \underset{\beta-{\rm
periodic}}{\int}\!\!\!\!\!\!{\mathcal{D}}\phi {\mathcal{D}}\pi \
 H_{I}(\phi,\pi) \exp\left(\int_{0}^{\beta} \! d \tau [i \dot{\phi}\pi -
H_0(\phi,\pi)] \right)/Z_0
\nonumber\\[2mm]
 &&\mbox{\hspace{12mm}}=\
 \mbox{Tr}\left\{ e^{-\beta \hat{H}_0} \ \!\left(\hat{\bar{H}}(\tau) - \hat{H}_0(\tau) \right) \right\}/Z_0 \nonumber\ =\    \mbox{Tr}\left\{ e^{-\beta \hat{H}_0} \ \!e^{-\tau \hat{H}_0}\left(\hat{\bar{H}}(0) - \hat{H}_0(0) \right)e^{\tau \hat{H}_0} \right\}/Z_0 \nonumber \\[2mm]
  &&\mbox{\hspace{12mm}}=\   \mbox{Tr}\left\{ e^{-\beta \hat{H}_0} \ \!
 \left(\hat{\bar{H}}(0) - \hat{H}_0(0) \right) \right\}/Z_0 \ \ = \ \langle H(0) - H_0(0)  \rangle_0 \, .
  \label{A.15.cf}
\end{eqnarray}
\end{widetext}
On the fourth line, we used the the cyclic property of the trace. With the help of~(\ref{A.15.cf}), we can rewrite~(\ref{A8.cv}) in the form
\begin{eqnarray}
F \ \leq \ F_0 \ + \ \langle H - H_0 \rangle_0 \, .
\label{A.10.jk}
\end{eqnarray}
This is the sought Bogoliubov--Feynman
inequality~\cite{Feynman}.  We can obtain the connection with Jarzynski's inequality~(\ref{2.aa})  in the following way. First, we define an interpolating operator $\hat{H}(\hat{\bar{\phi}}(0),\hat{\bar{\pi}}(0), \lambda)$ with $\lambda \in [0,1]$, where
\begin{eqnarray}
\hat{H}(\hat{\bar{\phi}}(0),\hat{\bar{\pi}}(0), \lambda = 1) \ &=& \ \hat{H}(\hat{\bar{\phi}}(0),\hat{\bar{\pi}}(0)) \nonumber \\[2mm] &\equiv &\ \hat{\bar{H}}(0)\, , \nonumber \\[2mm]
\hat{H}(\hat{\bar{\phi}}(0),\hat{\bar{\pi}}(0), \lambda = 0) \ &=& \ \hat{H}_0(\phi(0),\pi(0)) \nonumber \\[2mm]
&\equiv& \ \hat{H}_0(0)\, .~~~~~~~
\end{eqnarray}
With this we can write the last line in~(\ref{A.15.cf}) as
\begin{eqnarray}
&&\mbox{\hspace{-10mm}}\langle H(0) - H_0(0) \rangle_0 \nonumber \\[2mm] &&= \  \int_0^1 d \lambda \ \!\left\langle \frac{\partial H(\phi(0),\pi(0), \lambda)}{\partial \lambda} \right\rangle_0\, .
\end{eqnarray}\\
By assuming that that $\lambda$ itself is parametrized by time from some interval  $[0,T]$, so that $\lambda(0) = 0$ and $\lambda(T) = 1$, then the previous equation can be cast in the form
\begin{eqnarray}
&&\mbox{\hspace{-10mm}}
\langle H(0) - H_0(0) \rangle_0 \nonumber \\[2mm] &&= \ \int_{0}^T  dt \ \! \dot{\lambda}(t)  \ \!\left\langle \frac{\partial H(\phi(0),\pi(0), \lambda)}{\partial \lambda} \right\rangle_0\, .
\label{A.13.cf}
\end{eqnarray}\\
At this stage we can recall that the fields $\phi(\boldsymbol{r}, \tau)$ and $\pi(\boldsymbol{r}, \tau)$ used so far are so-called Euclidean fields. For the sake of explicitness we might denote them with a subindex ``$E\ \!$''. They are related to fields that depend on a real time $t$ (rather than inverse temperature $\tau$) as~\cite{Landsman}
\begin{eqnarray}
\phi_E(\boldsymbol{r}, it)  \ \!= \ \!  \phi(\boldsymbol{r}, t) \;\;\; \mbox{or} \;\;\;\;  \phi_E(\boldsymbol{r}, \tau)  \ \!= \ \!  \phi(\boldsymbol{r}, -i\tau)\, ,~~~
\end{eqnarray}
with $\phi_E(\boldsymbol{r}, 0) = \phi(\boldsymbol{r}, 0)$.  Similarly relations hold for the field $\pi$ and its euclidean version $\pi_E$.

Now, using the cyclic property of the trace and the fact that
\begin{eqnarray}
&&e^{it\hat{H}_0} \hat{\bar{\phi}}(\boldsymbol{r}, 0) e^{-it\hat{H}_0} \ = \ \hat{\bar{\phi}}(\boldsymbol{r}, t)\, , \nonumber \\[2mm] &&e^{it\hat{H}_0} \hat{\bar{\pi}}(\boldsymbol{r}, 0) e^{-it\hat{H}_0} \ = \ \hat{\bar{\pi}}(\boldsymbol{r}, t)\, ,
\end{eqnarray}
we can write~(\ref{A.13.cf}) as
\begin{widetext}
\begin{eqnarray}
\langle H(0) - H_0(0) \rangle_0  \ = \ \int_{0}^T  dt \ \! \dot{\lambda}(t)  \ \!\left\langle \frac{\partial H(\phi(t),\pi(t), \lambda(t))}{\partial \lambda} \right\rangle_0 \ =   \ \!\left\langle\int_{0}^T  dt \ \! \dot{\lambda}(t)  \frac{\partial H(\phi(t),\pi(t), \lambda(t))}{\partial \lambda} \right\rangle_0  \ = \   \left\langle W \right\rangle_0\, .~~~~~
\end{eqnarray}
\end{widetext}
Hence~(\ref{A.10.jk}) implies that
\begin{eqnarray}
\Delta F \ \leq \  \langle W\rangle_0\, ,
\end{eqnarray}
which is yet another variant of JI. Note that in the derivation we have only assumed that the initial and final states are in thermal equilibrium within a heat bath of equal temperature $1/\beta$. The process between initial and finite time could be arbitrarily far from equilibrium, since the control parameter $\lambda(t)$ could in principle have arbitrary time dependence (provided the constraints $\lambda(0) = 0$ and $\lambda(T) = 1$ are satisfied).
Note in particular, that the average is only with respect to the initial-time thermal density matrix. This differs from the conventional JI where the average is over an ensemble of all possible phase-space trajectories obtained by sampling initial conditions from a canonical
ensemble.

\vspace{3mm}
A few comments are now in order:

\begin{enumerate}
\item The result obtained can easily be generalized also to situations when chemical potential is present. This can be seen from the fact that\\[-1mm]
\begin{eqnarray}
~~~~Z \ &=& \  \mbox{Tr}\left(e^{\beta \sum_a \mu_a \hat{N}_a  } e^{-\beta \hat{H}}   \right) \ = \ \mbox{Tr}\left(e^{-\beta \ \! \hat{\bf{H}}} \right)\nonumber \\[2mm]
&=&  \ \mbox{Tr}\left(e^{-\beta \ \! \hat{\bf{H}}_0}    \ \! e^{\beta \hat{H}_0} e^{-\beta \hat{H}}\right) \nonumber \\[2mm] &=& \ Z_0 \left\langle   \hat{\Lambda}(-i\beta)  \right\rangle_0 ,~~~~~~~~
\end{eqnarray}
where the bold Hamiltonians also include the particle species number operators $N_a$ with the ensuing chemical potentials $\mu_a$.  The functional integral representations are the same
as~(\ref{app.1.1aa})-(\ref{app.1.1a}) but now with bold Hamiltonians and the grand potential $\Omega$ instead of Helmholtz free energy $F$. The proof of the BF inequality thus follows the same chain of reasoning as before.
The corresponding generalization of JI will read
\begin{eqnarray}
~~~~~~~~~\Delta \Omega \ \leq \ \langle W \rangle_0\, .
\label{IV.113.gh}
\end{eqnarray}
Here $\langle \cdots \rangle_{0}$ denotes an average taken with respect to the initial-time density matrix associated with the Hamiltonian $\hat{\mathbf{H}}_{0}$. The work $W$ in Eq.~(\ref{IV.113.gh}) represents the total work performed on the system, including both mechanical and chemical contributions.

\item Although the derivation in this section was carried out only for scalar fields (i.e., spinless particles), the same steps can be straightforwardly extended to spin-$\tfrac{1}{2}$ fields using Berezin functional integral over Grassmann variables, as well as to Yang--Mills gauge fields (i.e., spin-1 particles).

\end{enumerate}

\section{Conclusions \label{conclusions}}

In this paper, we derived several variants of Jarzynski's inequality under different thermodynamic constraints, including conditions of constant temperature and pressure and of constant temperature and volume. We  also addressed the connection between JI and the maximum work theorem. Our analysis spans classical and quantum field–theoretical regimes with a particular focus on dissipative chemical systems (both in the linear-response regime and away
from linear thermodynamics), and to the mechanisms by which the JI emerges in these frameworks. In all cases considered, we find that the average work performed under the respective constraints is bounded from below by the corresponding free-energy change. These results demonstrate the broad validity of JI, which often goes beyond the conventional regimes in which JE is typically phrased (and applied), and across a wide range of thermodynamic settings. The latter further suggest that more general proofs of JE may exist, establishing its validity in a broader class of settings than those in which it is conventionally assumed to hold.


Finally, we have demonstrated the usefulness of functional-integral techniques for deriving JI within a QFT framework. We expect that this approach can be further extended beyond averaging with respect to the initial-time Hamiltonian by employing the Feynman--Vernon influence functional~\cite{Feynman2} and the Schwinger--Keldysh closed-time-path formalism~\cite{Calzeta,Landsman}. Work along these lines is currently in progress.




\vspace{6pt}

\acknowledgments{
Both D.R.C. and P.J. were supported by the Czech Science Foundation Grant (GA\v{C}R), Grant No. 25-18105S. D.R.C. was also, in part, supported by the Grant Agency of the Czech Technical University in Prague, Grant No. SGS25/163/OHK4/3T/14.
}






\appendix



\section{~~Structure of Legendre transforms for irreversible processes \label{A1cc}}

For the sake of consistency and notational unity, we will now briefly review the key aspects of Legendre transforms and the ensuing inequalities in irreversible processes that are needed in the bulk of the paper [namely in Secs.~\ref{section 3.a} and~\ref{chapter 4.a}]. It should be noted that the Legendre-transformed structure and the resulting potentials are not restricted to equilibrium or the linear response regime. In fact, they remain applicable even away from linear thermodynamics (including systems far from equilibrium) as long as local equilibrium, time-scale separation, and convexity of the underlying thermodynamic potentials are preserved. In such case, nonlinearity affects constitutive relations but does not invalidate the Legendre structure itself~\cite{prigogine1955thermodynamics,GM}.

In irreversible systems, the starting point is the second law based on the production of entropy, which states that
\begin{eqnarray}
&&dS_i \ > \  0\, , \quad \text{irreversible evolution}\, , \label{A.1.dk}\\[2mm]
&&dS_i \ = \  0\, , \quad \text{reversible evolution}\, .
\end{eqnarray}
In particular, Eq.~(\ref{3.3.aa}) can be written in the form
\begin{eqnarray}
&&dS_s \ = \  \frac{\delta Q}{T} \ + \  dS_i\, , \nonumber \\[2mm]
&&\delta Q  \ = \ TdS_s \ - \  TdS_i\, .
\label{A.3}
\end{eqnarray}
Combining equation~(\ref{A.3}) with the first law of thermodynamics then leads to
\begin{eqnarray}
dU \ &=& \  TdS_s \ - \  TdS_i  \  + \ \delta W \nonumber \\[2mm] &=& \ TdS_s \ - \  TdS_i   - \ pdV \ + \ \delta W_{\rm{nm}} \, ,
\end{eqnarray}
where $\delta W$ is the total work done on the system and $\delta W_{\rm{nm}}$ is the non-mechanical work done on the system.
With the help of the Legendre transform, we can convert the internal energy $U$ into another thermodynamic potential by changing the set of independent variables, namely
\begin{eqnarray}
&&d(U \ - \  TS_s) \ = \  TdS_s \ - \  TdS_i \ - \ pdV   \nonumber \\[2mm] &&\mbox{\hspace{28mm}}+ \ \delta W_{\rm{nm}} \ - \ d(TS_s)\, ,\nonumber \\[2mm]
&&\Rightarrow\ \ dF \ = \  -S_sdT \ - \  TdS_i \ - \ pdV  \ + \ \delta W_{\rm{nm}} \, ,~~~~~~~~
\label{A.5.hj}
\end{eqnarray}
and similarly
\begin{eqnarray}
&&\mbox{\hspace{-4mm}}d(U \ + \  pV \ - \  TS_s) \ = \  TdS_s \ - \  TdS_i \ - \ pdV  \nonumber \\[2mm] &&\mbox{\hspace{29mm}}+ \ \delta W_{\rm{nm}}  \ - \ d(TS_s) \ + \ d(pV)\, , \nonumber  \\[2mm]
&&\mbox{\hspace{-4mm}}\Rightarrow\ \ dG \ = \  -S_sdT \ + \  Vdp \ - \  TdS_i  \ + \ \delta W_{\rm{nm}} \, .
\label{A.6.cc}
\end{eqnarray}
If the variables describing the system are $T$ and $V$, the corresponding potential is the Helmholtz free energy $F$. When we fix only the temperature $T$, we obtain
\begin{eqnarray}
dF_T \ &=& \   -pdV   \ + \ \delta W_{\rm{nm}} \ - \  TdS_i \nonumber \\[2mm] & \leq & \  -pdV   \ + \ \delta W_{\rm{nm}} \ = \ \delta W\, .
\label{A.16.bv}
\end{eqnarray}
The equality holds only for reversible processes.  This relation  is important in Sec.~\ref{section 3.a}.


If the variables describing the system are $T$ and $p$, the corresponding potential is the Gibbs free energy $G$. By imposing the constraint that $T$ and $p$ are constant, then from~(\ref{A.6.cc}) and~(\ref{A.1.dk}) follows that
\begin{eqnarray}
dG_{T,p} \ = \ - \  TdS_i  \ + \ \delta W_{\rm{nm}}  \ \leq \ \delta W_{\rm{nm}} \, ,
\label{A.17.hj}
\end{eqnarray}
with equality if and only if the process is reversible.
We might note in passing that this implies that [cf.~relation~(\ref{eq:myequation11})]
\begin{eqnarray}
dG_{T,p}  \ \leq \ -A d\xi\;\;\;\; \Rightarrow\;\;\; \; \left(\frac{\partial G}{\partial \xi}  \right)_{T,p}  \ \leq \ -A\, .
\label{A.9.kl}
\end{eqnarray}
%
Both relations~(\ref{A.16.bv}) and~(\ref{A.17.hj})  are employed in Secs.~\ref{section 3.a} and~\ref{chapter 4.a}.


%







\end{document}